*Patterning active materials with addressable soft interfaces*

Pau Guillamat, Jordi Ignés-Mullol, and Francesc Sagués*

Departament de Química Física and Institute of Nanoscience and Nanotechnology (IN2UB), Universitat de Barcelona. Martí i Franquès 1, 08028 Barcelona. Catalonia. Spain.

**Motor-proteins are responsible for transport inside cells. Harnessing their activity is key towards developing new nano-technologies [1], or functional biomaterials[2]. Cytoskeleton-like networks, recently tailored in vitro[3-6], result from the self-assembly of subcellular autonomous units. Taming this biological activity bottom-up may thus require molecular level alterations compromising protein integrity. Taking a top-down perspective, here we prove that the seemingly chaotic flows of a tubulin-kinesin active gel can be forced to adopt well-defined spatial directions by tuning the anisotropic viscosity of a contacting lamellar oil. Different configurations of the active material are realized, when the passive oil is either unforced or commanded by a magnetic field. The inherent instability of the extensile active fluid[7,8] is thus spatially regularized, leading to organized flow patterns, endowed with characteristic length and time scales. Our finding paves the way for designing hybrid active/passive systems where ATP-driven dynamics can be externally conditioned.**

Cytoskeleton reconstitutions are structured soft systems displaying characteristics of self-assembled colloidal dispersions, but, distinctively, they are active [3-6,9-15]. Under suitably constrained conditions, and with a steady supply of ATP, these new types of materials feature emergent modes of self-organization that arise from the collective behavior of individual biomolecules. Besides posing fascinating questions, related to the permanent and intrinsic non-equilibrium nature of these systems[16,17], the fact that they constitute *in vitro* models for the intra-cellular milieu suggests a potential for the development of new responsive biomaterials. For instance, one could envision to externally commanding the preferred flow lines, deformation directions, or intrinsic entanglements of these active gels, although a successful strategy in this direction has yet to be demonstrated. The complexity and specificity of the involved components seems to preclude an *a priori* process of protein engineering. It seems thus advisable to devise a control strategy with the potential of being exerted on any viable active gel. Actuating by means of either an electric or a magnetic field is unlikely, given the high ionic strength of these preparations, which screens electric responses, and the low magnetic susceptibilities that make them insensitive to modest magnetic forces. On the other hand, one could resort to the confinement of the active material in patterned microfluidic channels, which will determine the direction and time scales of its spontaneous active deformations. This procedure has the disadvantage of requiring complex biochemical functionalization of the inner surfaces to render the substrates biocompatible but, even if successful, this strategy will be neither versatile nor *in situ* reconfigurable. Here, we demonstrate a radically different approach by interfacing the aqueous active gel with an oily component that features smectic (lamellar) liquid-crystalline order[18]. Such compounds have been known for a long time to easily align in the presence of modest electromagnetic fields, and to have dramatic anisotropies in their shear viscosities, which we employ here to tame the seemingly chaotic activity of protein gels.

The chosen active material is based on the self-assembly of tubulin, from protein monomers all the way up to filamentary bundles of micron-sized stabilized microtubules[4,19]. The latter are cross-linked and sheared by clusters of ATP-fueled kinesin motors, which are directed towards the plus ends of the microtubules. Inter-filament sliding thus occurs in bundles containing microtubules of opposite polarity (Figure S1 in Supplementary Information). This mixture self-assembles into an extensile active gel[20-22], continuously rebuilt following bundle reconstitution, and permanently permeated by streaming flows. An alternative preparation, more suited to our purposes, consists in depleting this bulk material towards a biocompatible soft and flat interface, where filaments continuously fold and adopt textures typical of a two-dimensional nematic[4]. This active layer appears punctuated by a steady number of continuously renovated microtubule-void regions that configure semi-integer defect areas. Although we focus on flat interfaces, the latter can be also curved, giving rise to interesting defect accommodation dynamics, and occasionally to intriguing deformation modes of the, in this way prepared, active vesicles[23]. In our case, a layer of active gel is confined between a biocompatible hydrophilic rigid surface (bottom) and a (top) volume of the hydrophobic oil octyl-cyanobiphenyl (8CB), which features liquid crystal behavior at temperatures compatible with protein activity. The oil/water interface is stabilized with a polyethylene glycol (PEG)-based triblock copolymer surfactant. Real time observation is performed using fluorescence, from tagged tubulin moieties, together with polarization and confocal reflection microscopies (See Experimental Methods in Supplementary Information).

To better appraise the role played by the contacting structured interface on the active material, the top passive liquid crystal is initially prepared in its nematic phase. On average, the 8CB mesogen molecules lay parallel to the oil/water interface, due to the influence of the polymeric surfactant, and perpendicular to the oil/air interface far from the active layer (Fig. 1a). The contacting active nematic features the well-known disordered swirling motion, characterized by the formation and annihilation of defects[24-27], accompanying the spontaneous folding of the bundled microtubules. Although the passive mesophase is anisotropic, shear viscosities along different directions are of the same order of magnitude, and do not trigger any observable alignment on the active nematic, which remains "well-mixed" at large-scales (Figs. 1b, 1c). However, when comparing with the original situation[4], where the contact is provided by a isotropic and much less viscous oil, we evidence remarkable differences relative to both the (number) density of proliferating defects, and the tracked velocities of the singularities. In short, the velocity decreases and the density of defects increases with the oil viscosity. This clearly points out to a marked effect of the interface on the active nematic, which arises from the need to accommodate the rheological properties of the latter to those of the passive contacting fluid[28,29]. This gives us a clear indication that the contact of the active and passive structured materials is robust, persistent, and stable enough to foresee stronger effects when the order of the passive phase is further upgraded to attain a lamellar (i.e. layered) disposition.

A temperature quench below 33.4 °C triggers the reversible transition of 8CB into the lamellar smectic-A phase. This has a dramatic impact on the interfacial rheology and on the dynamics of the active nematic. Layers in the smectic-A phase are perpendicular to the oil molecules, which remain parallel to the oil/water interface. Free-energy minimization constraints, related to the boundary conditions at the interfaces, result in the formation of the so-called toroidal focal conics[18]. These are polydisperse circular domains where radially-oriented mesogen molecules arrange concentrically at the oil/water interface (Fig. 1d). Flow perpendicular to the smectic layers is severely hindered and, contrarily, is highly facilitated along them. As a result, the local interfacial shear stress experienced by the active material is markedly anisotropic. Stretching of the bundled filaments thus occurs preferably along circular trajectories centered with the focal conics, which organize the active nematic in confined rotating mills (Fig. 1e and Video S1 in Supplementary Information). Flows can be traced following the motion of the +1/2 comet-like defects, and become fully apparent from time-averaged snapshots of the active pattern. Dark spots, where filament occupancy is lowest, appear then in register with the center of large focal conics in the smectic-A oil (Fig. 1f). Increasing the temperature, the smectic phase of the passive liquid crystal melts down and, as expected, the rotating mills are completely and immediately dismantled, recovering the typical textures of the active nematic (see Video S2 in Supplementary Information).

Filamentary bundles of the active material have an intrinsic maximum curvature that cannot be exceeded. As a consequence, only focal conics above a threshold diameter, which we have estimated at around 40 μm in the experiment shown in Fig. 1, are able to coordinate the flow of the underlying active material. From a topological perspective, confined rotating filaments organize a topological singularity of charge S=+1, which corresponds to a full rotation around the domain center (Fig. 2a). The semi-integer defects, which accompany rotating filaments, are permanently renovated, even their number may change to a small extend, assembling and disassembling continuously the core structure of the vortex (see Video S3 in Supplementary Information). At all times, however, a balance is established such that the arithmetic sum of topological charges inside a single domain adds up to one (Fig. 2b). This collective precession is accompanied by a periodic modulation of the velocity of the active flows (Fig. S2 in Supplementary Information).

A tighter control of the active flow is achieved by directly actuating on the contacting passive oil. To this purpose, we take advantage of its large (positive) diamagnetic anisotropy that enables to align a macroscopic volume of the material with uniform magnetic fields of the order of a few kG, easily attainable with a permanent magnet array (see Experimental Methods in Supplementary Information). By a slow temperature quench of 8CB from the nematic to the lamellar (smectic-A) phase in presence of a 4 kG magnetic field parallel to the oil/water interface, we induce the formation of a layer of oil molecules uniformly aligned with the magnetic field. In this situation, the lamellar planes are oriented perpendicularly both to the interface and to the magnetic field (Fig. 3a). It is well-known in the literature that this *bookshelf* geometry, robustly kept after removal of the magnetic field, results in a liquid that flows easily when sheared along the lamellar planes, but that responds as a solid to stresses exerted in the orthogonal direction[18]. Polarizing optical microscopy confirms the formation of this aligned smectic-A layer (Fig. 3a) that includes dislocations in the aligned lamellar planes, which propagate into the bulk of the material with the so-called parabolic focal conic domains. In contact with this interface, the turbulent active nematic experiences markedly anisotropic shear stresses, which result in a rapid rearrangement of the flow pattern that consists now in bands of uniform width, perpendicular to the magnetic field (Figs. 3b, 3c and Video S4 in Supplementary Information). More precisely, stripes of densely stacked microtubules that appear bright in the fluorescence images, appear alternated with dark lanes of moving defects, along which antiparallel flow velocities are easily recognized, and clearly visualized using particle tracers (see Video S5 in Supplementary Information). Negative defects are, in this situation, more difficult to localize, although the global topological requirement mentioned earlier applies here to secure a zero total charge. As before, this organized structure is immediately and totally reset, to recover the standard well-mixed configuration of the active material, when reversing the temperature quench back to the nematic phase of the passive oil (see Video S6 in Supplementary Information). Moreover, by previous transiting back and forth through the passive nematic 8CB phase, the orientation of the active pattern can be readily and reversibly in plane-rotated by redirecting the magnetic field at will (see Video S6 in Supplementary Information).

Within stripes, bundles of microtubules are packed in parallel arrangements oblique to the lanes of moving defects (Fig.4a and Fig. S3 in Supplementary Information). This microtubule disposition, although apparently motionless, is prone to suffer the bending instability intrinsic to any extensible active material [7,8]. Following such instability, pairs of complementary half-integer defects are created and either they annihilate in pairs or incorporate into opposite preexisting lanes (Fig. 4b and Video S7 in Supplementary Information). The transit across lanes can be visualized using particle flow tracers (Video S5 in Supplementary Information). Once defects escape from their birth areas, the striped regions are restored with minimal rearrangements. Such apparently vulnerable episodes of the active material extend coherently over regions of arbitrary extension but always commensurate with the stripe width, and occur periodically with a striking regularity (see Video S7 in Supplementary Information). Actually, when tracing the pattern of active flows, a neat oscillation of the velocity, averaged over the entire observed area, is readily captured (Fig. 4c). This characteristic frequency increases with the activity of the sample, here quantified in terms of the concentration of ATP (Fig. 4c). We conjecture that this kind of periodic "avalanches" arise from the intrinsic dynamics of the sheared microtubules, and provide a breakdown mechanism that the active material has at its disposition to repeatedly release the extensile tensional stress accumulated in the stripes of the active material. The striped pattern also defines a characteristic wavelength, which depends on the activity of the sample (Fig. 4d). This characteristic spatial periodicity is likely to be associated to a typical length scale in the active material that has been predicted in theoretical studies to govern the decay of the flow vorticities and the director fields for unconfined active nematics[30].

In summary, our work demonstrates that the disordered flow patterns of an active two-dimensional nematic can be controlled to follow preassigned directions. This is achieved by means of the anisotropic shear stress exerted at the interface by a liquid crystal layer whose orientation is externally commanded with a magnetic field. The ordered active material clearly reveals its characteristic time and length scales, which arise from non-equilibrium biochemical processes at the molecular level. The versatility, reversibility and robustness of such strategy should be considered as a proof of concept for the taming of active subcellular materials.

**Acknowledgements**

The authors are indebted to Z. Dogic and S. DeCamp (Brandeis University) for their assistance in the preparation of the active gel. We thank M. Pons (Universitat de Barcelona) for his assistance in the expression motor proteins. Funding has been provided by MINECO (project FIS 2013-41144P). P.G. acknowledges funding from Generalitat de Catalunya through a FI-DGR PhD Fellowship.

**Author contributions**

All authors conceived the experiments. J.I.-M. designed and built the experimental setup. P.G. performed the experiments and analyzed experimental data. F.S. wrote the manuscript. P.G. and J.I.-M. prepared the figures. All authors discussed, and revised the manuscript.

**Competing financial interests**

The authors declare no competing financial interests.

**Figure captions**

**Figure 1. Self-assembly of the active layer in contact with the unforced liquid crystal.** The active material is in contact either with the nematic (**a-c**) or the smectic-A phase (**d-f**) of the oily mesogen. Reversible transition between both states is observed at 33.4 °C. Confocal reflection micrographs of the interface reveal the featureless nematic 8CB phase, where molecules lay, on average, parallel to the oil/water interface, with short-range positional but long-range orientational correlations (**a**). Fluorescence confocal microscopy imaging of the active nematic depleted towards the low viscosity weakly anisotropic oil (**b**) reveals the characteristic semi-integer folds (highlighted in **b**) that dynamically form, annihilate, and move in a turbulent fashion, with no temporal coherence, as evidenced by the time average shown in (**c**). A temperature quench into the lamellar smectic-A phase of 8CB results in the formation of the well-known toroidal focal conic domains, as revealed by confocal reflection imaging (**d**). Inside these domains, radially-oriented molecules organize in concentric circular layers. The resulting anisotropy in shear viscosity triggers the reorganization of the active nematic underneath in confined rotating mills as evidenced both from instantaneous (**e**) and time-averaged fluorescence confocal microscopy snapshots (**f**). Micrographs are 244 µm wide. (See Video S1 in Supplementary Section).

**Figure 2. Topology of the active nematic when constrained to circular flows. a** Fluorescence micrograph of a vortex of the active nematic, which has formed in contact with a toroidal focal conic domain in the 8CB film, and velocity plot (right). Green arrows indicate the direction and magnitude of velocity at each point. **b** Micrographs with two-channel overlay of the confocal fluorescence (green) and confocal reflection (grayscale) microscopy displaying different arrangements of the active nematic filaments constrained by underlying toroidal focal conic domains. Topological defect configurations satisfy the constraint that the net charge must be S=+1 inside a circular domain. In the shown cases: (left), S = 2 × (+½); (center) S = 3 × (+½) - ½; (right) S = 4 × (+½) - 1. Scale bar, 50 µm. (See Video S3 in Supplementary Information).

**Figure 3. Active layer forced by the magnetically aligned liquid crystal. a** Polarized optical micrograph of the passive oil in the lamellar phase after being aligned with a magnetic field along the X direction, thus adopting the so-called *bookshelf* geometry. Layers of 8CB molecules organize forming elongated parabolic focal conic domains that include dislocations, as shown in the sketch. **b** Fluorescence confocal micrographs of the active nematic depleted towards

the water/oil interface. The otherwise random proliferation and dynamics of filament folding is organized, by contact with the aligned hydrophobic mesogen, in the form of stripes where active filaments are dense, intercalated by lanes, where defect cores organize antiparallel flow directions. A time-averaged snapshot (**c**) clearly illustrates this anisotropic self-assembly, with arrows depicting flow directions. Micrographs are 375 µm wide. (See Video S4 in Supplementary Information).

**Figure 4. Oscillations of the aligned active nematic. a** The dynamically self-assembled state where active filaments are indirectly aligned by a magnetic field is metastable. Adjacent defect lanes are linked by quasi-stationary stripes where filaments are densely packed. The arrowheads indicate the direction of defect motion. Scale bar, 30 µm. **b** Parallel extensile filament bundles are prone to bend distortions that lead to the proliferation of topological defects (see Video S7 in Supplementary Information). Blue circles and red triangles point the position of +1/2 and -1/2 defects, respectively. Fluorescence micrographs (left) are shown alongside sketches (right). **c** The average interfacial velocity oscillates due to the periodic disruption of the aligned stripes of bundled filaments. Higher ATP concentration, which leads to increased activity, results in higher velocities and faster oscillatory dynamics. **d** Time-averaged fluorescence microscopy snapshots and plot of the inter-lane spacing as a function of [ATP]. Besides stablishing a characteristic time scale, activity also controls the distance between defect lanes. ATP concentration is 1400, 700, 470, 280 and 140 µM from the top to the bottom. Images are 550 µm wide.

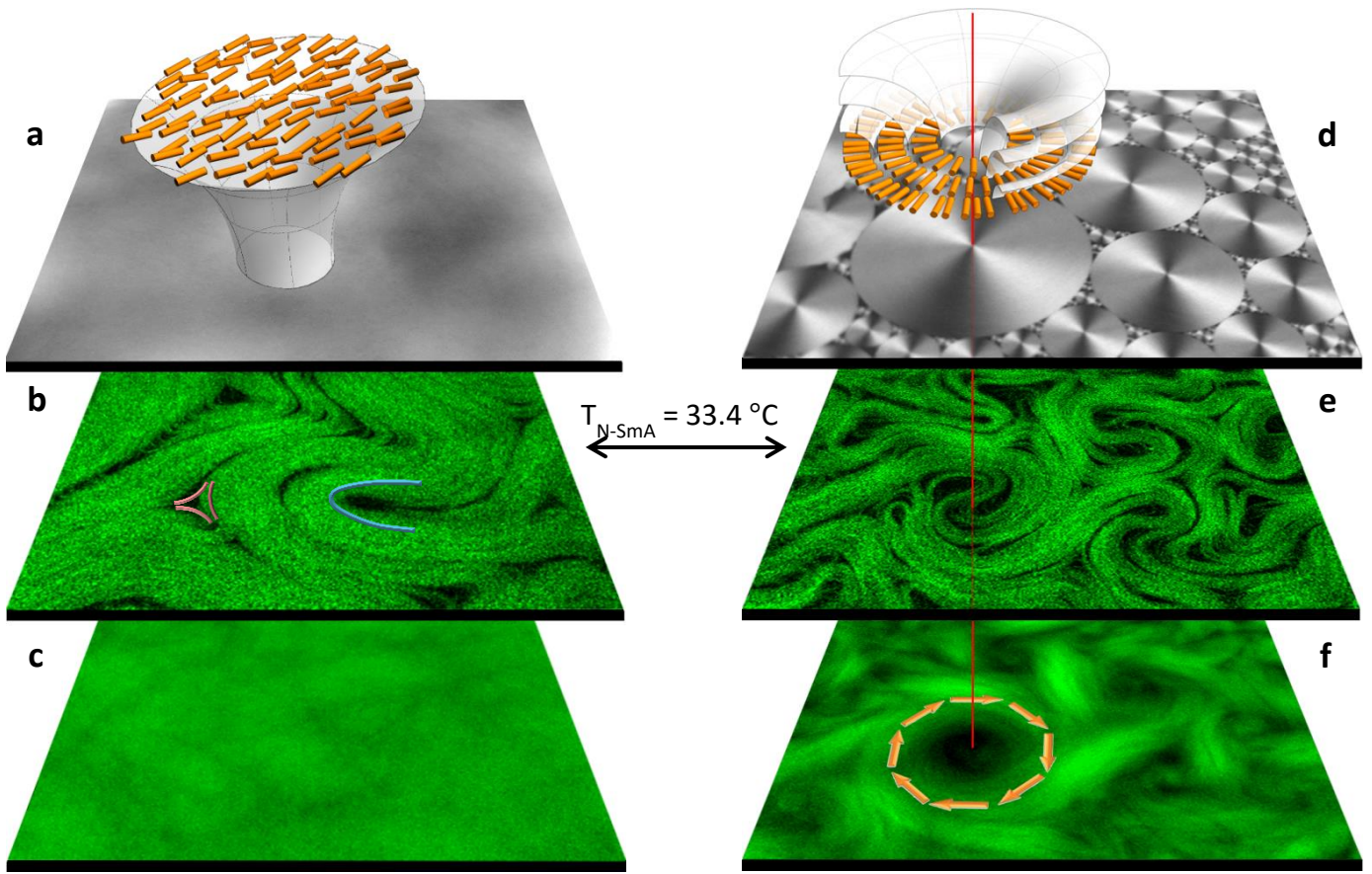

Figure 1

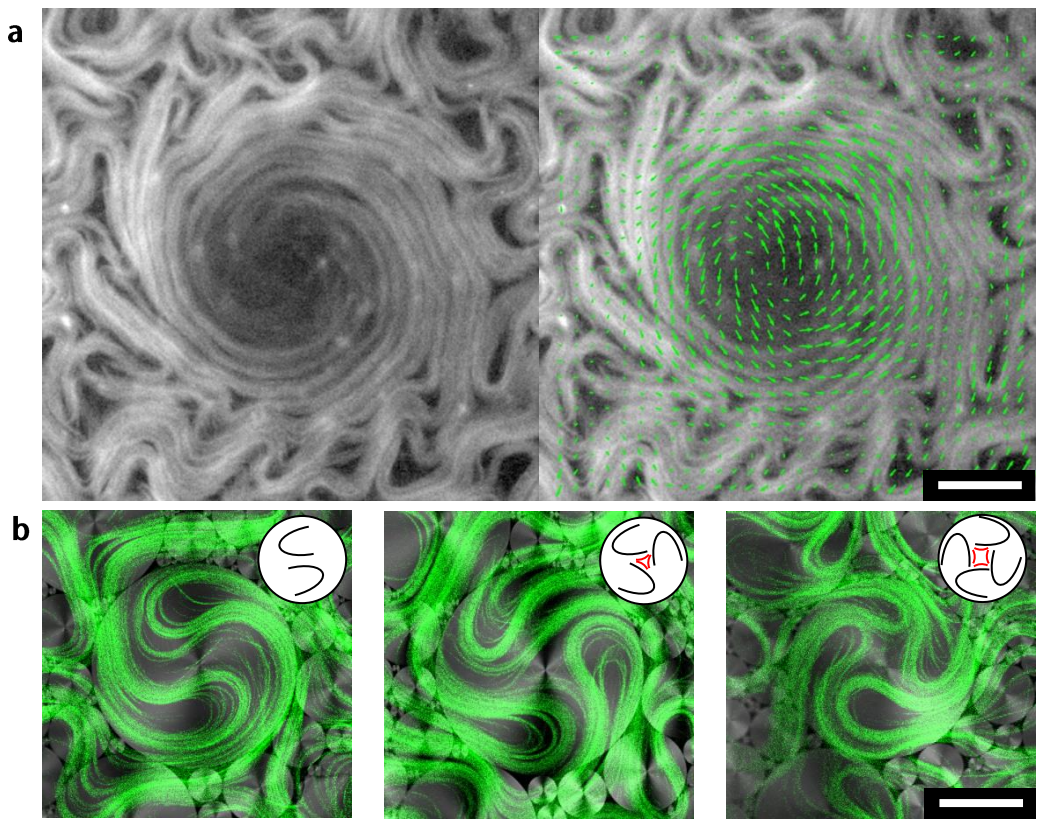

Figure 2

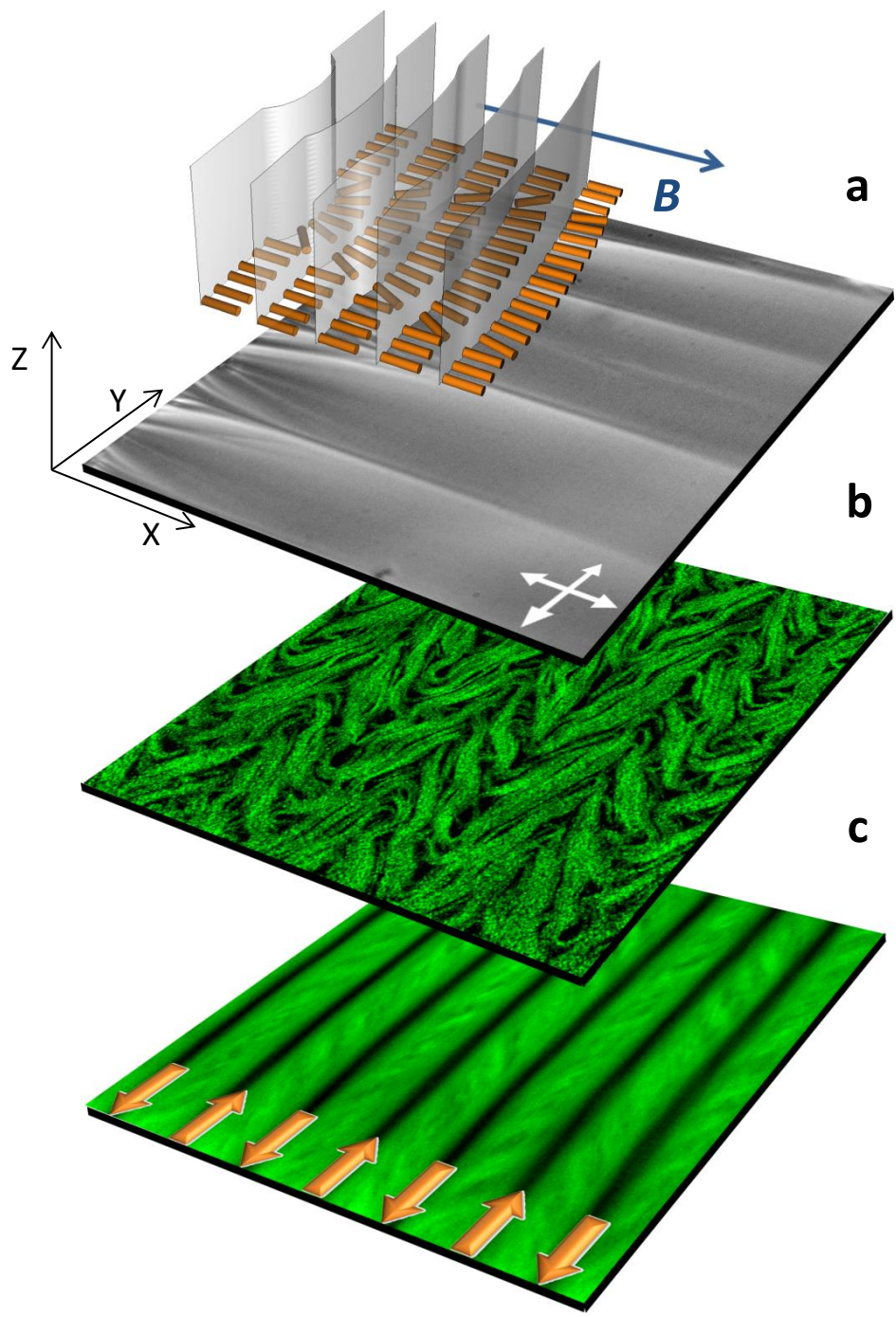

Figure 3

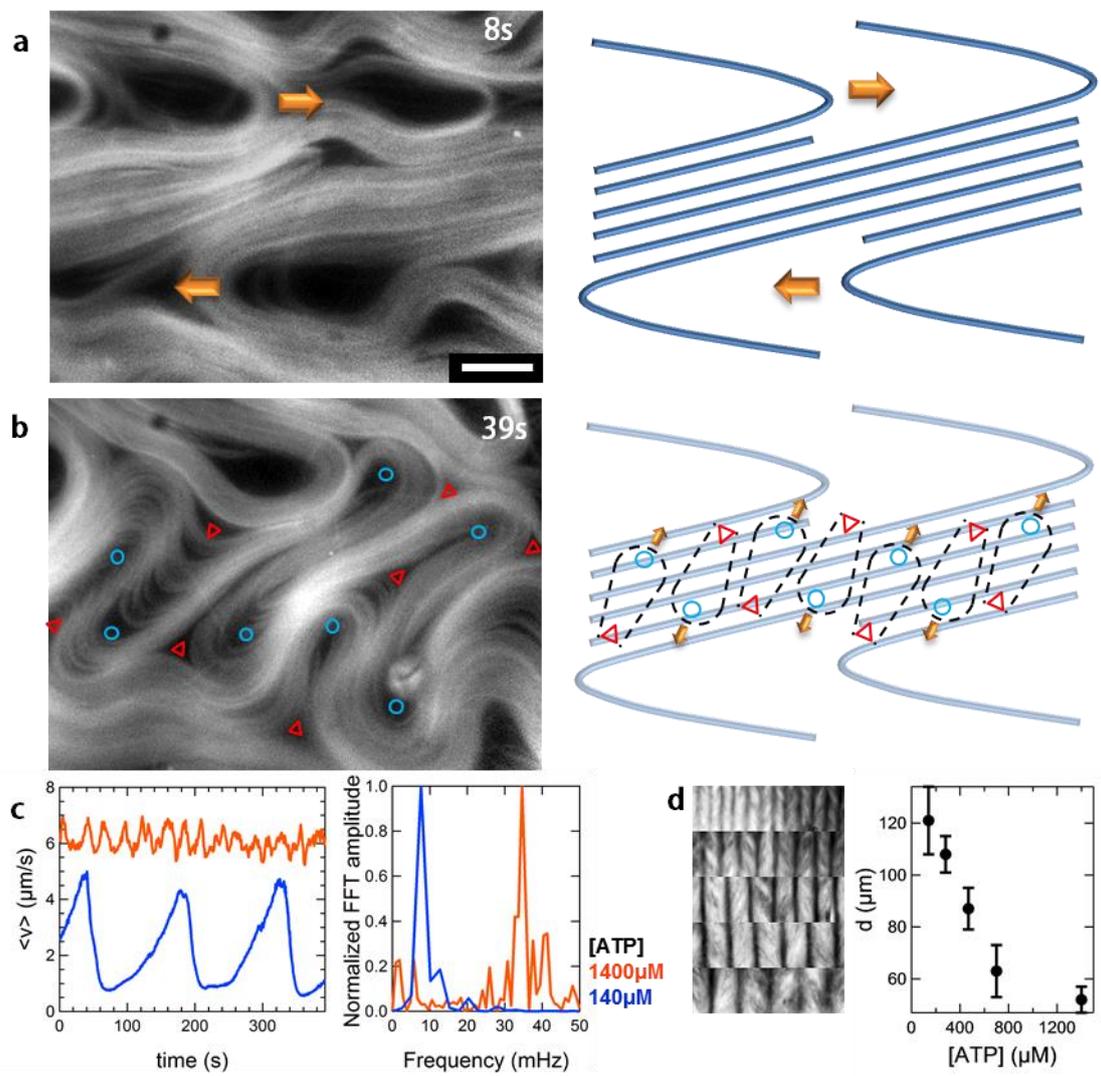

Figure 4

# Patterning Active Materials with Adressable Soft Interfaces

P.Guillamat, J.Ignés-Mullol, F.Sagués

Departament de Química Física and Institute of Nanoscience and Nanotechnology (IN2UB), Universitat de Barcelona. Martí i Franquès 1, 08028 Barcelona. Catalonia. Spain.

**Table of contents**

1. Experimental Methods
2. Supplementary Figures
3. Supplementary Videos
4. References

# 1. Experimental methods

## 1.1. Thermotropic liquid crystal

4-cyano-4'-octylbiphenil (8CB) is a thermotropic liquid crystal between 21.4 and 40.4ºC, featuring a transition from the smectic-A (SmA) to the nematic (N) phase at 33.4ºC. Like most usual organic mesogens, 8CB is diamagnetic, and exhibits positive diamagnetic anisotropy ($\chi_a = \sim 10^{-6}$) (*1, 2*). Therefore, a magnetic field of the order of 1 kG will exert a torque that is able to align a layer of 8CB confined between anchoring boundaries.

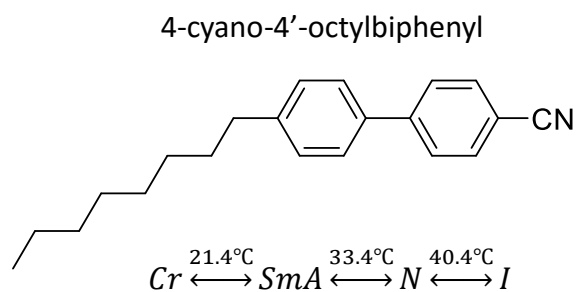

$$Cr \xleftrightarrow{21.4°C} SmA \xleftrightarrow{33.4°C} N \xleftrightarrow{40.4°C} I$$

## 1.2. Tubulin-based active gel

### 1.2.1. Polymerization of microtubules

Heterodimeric (α,β)-tubulin from bovine brain (a gift from Z. Dogic's group in Brandeis University) is incubated at 37ºC for 30 min in an M2B buffer (80 mM PIPES, 1 mM EGTA, 2 mM MgCl$_2$) supplemented with the reducing agent dithiothrethiol (DTT, Sigma, 43815) and with Guanosine-5'-[(α,β)-methyleno]triphosphate (GMPCPP, Jena Biosciences, NU-405), a slowly hydrolysable analogue of the biological nucleotide Guanosine-5'-triphosphate (GTP) that completely suppresses dynamic instability of the polymerized tubulin microtubules (MTs) (*3*). GMPCPP enhances spontaneous nucleation of MTs (*3*) obtaining high-density suspensions of short MTs (1-2 µm). For fluorescence microscopy, a fraction of the tubulin (3%) has been fluorescently labelled with Alexa-647.

### 1.2.2. Kinesin expression

*Drosophila Melanogaster* heavy chain kinesin-1 K401-BCCP-6His (truncated at residue 401, fused to biotin carboxyl carrier protein (BCCP) and labelled with 6 Histidine tags) has been expressed in *E.coli* using the plasmid WC2 from Gelles Lab (Brandeis University), and purified with a nickel column (*4*). After dialysis against 500 mM Imidazole buffer, kinesin concentration is estimated by means of absorption spectroscopy and stored at the desired concentration in a 60% sucrose solution at -80 ºC. (*3*)

### 1.2.3. Preparation of molecular motor clusters

The biotin-streptavidin pair presents one of the strongest and more specific noncovalent interactions known in biochemistry (*5*). In our experiments, biotinylated kinesin motor protein and tetrameric streptavidin (Invitrogen, 43-4301) are incubated on ice for 30 minutes at the specific stoichiometric ratio 2:1 in order to obtain kinesin-streptavidin motor clusters.

*1.2.4. Assembly of MT-based active gel*

MTs are mixed with a solution containing a non-adsorbing polymeric depleting agent (polyethylene glycol, PEG, 20kDa, Sigma, 95172) that bundle the filaments together, molecular motor clusters that act as cross-linkers, and ATP (Sigma, A2383) that drives the activity of the gel (Fig. S1). In order to maintain a constant concentration of ATP during the experiments, an enzymatic ATP-regenerator system is used, consisting on Phosphoenol pyruvate (PEP, Sigma, P7127) that fuels Pyruvate kinase/lactate dehydrogenase (PK/LDH, Invitrogen, 434301) to convert ADP back into ATP. Several anti-oxidant components are also included in the solution in order to avoid protein denaturation and to minimize photobleaching during characterization by means of fluorescence microscopy. The PEG-based triblock copolymer surfactant Pluronic F-127 (Sigma, P-2443) is also added at 2 %w/w (final concentration) in order to procure a biocompatible water/oil interface in subsequent steps.

## 1.3. Experimental Setup

A block of Poly-dimethylsiloxane (PDMS) with a cylindrical pool of diameter 5 mm and height 8 mm is manufactured using a custom mold. The block is glued onto a bioinert and superhydrophilic polyacrylamide-coated glass, which is prepared following ref.

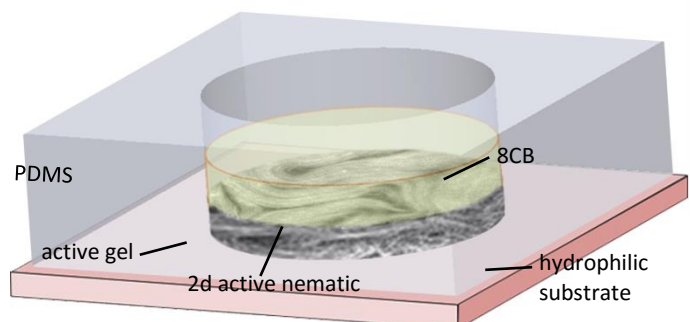

(*6*). In brief, clean and activated glass is first silanized with an acidified ethanolic solution of 3-(Trimethoxysilyl)propylmethacrylate (Sigma,440159), which will act as polymerization seed. The silanized substrates are subsequently immersed in a solution of acrylamide monomers (for at least 2 h) in the presence of the initiator ammonium persulphate (Sigma, A3678) and N,N,N',N'-Tetramethylethylenediamine (TEMED, Sigma, T7024), which catalyzes both initiation and polymerization of acrylamide.

The pool is first filled with 50 µL of 4-cyano-4'-octylbiphenil (8CB, Synthon, ST01422) and, subsequently, 1 µL of the active gel is injected at the bottom of the oil. The system is heated up to 35 ºC in a homemade oven in order to promote transition to the less viscous nematic phase of the mesogen, which facilitates the spreading of the active gel onto the acrylamide substrate. After several minutes at room temperature, the active nematic is spontaneously formed at the flat 8CB/water interface. Unlike the conventional flow cells, in which a layer of the active gel is confined in a thin gap between two glass plates, this setup enables us to use high viscosity oils to prepare the interface.

## 1.4. Temperature control and magnetic field application

Polarizing and fluorescence microscopy is carried out in a custom optical setup, built around a custom-made permanent magnet assembly that provides homogeneous planar magnetic field in a region much larger than the field of view, with a maximum strength of 4.4 kG. The magnet is built as a Halbach cylindrical array(*7*) consisting on eight identical N52-grade NdFeB cubic magnets (cube size = 25.4 mm, K&J Magnetics BX0X0X0-N52). The magnets are assembled, with the suitable geometric arrangement and magnetic moment orientation, using a 3D-printed Poly-lactic acid 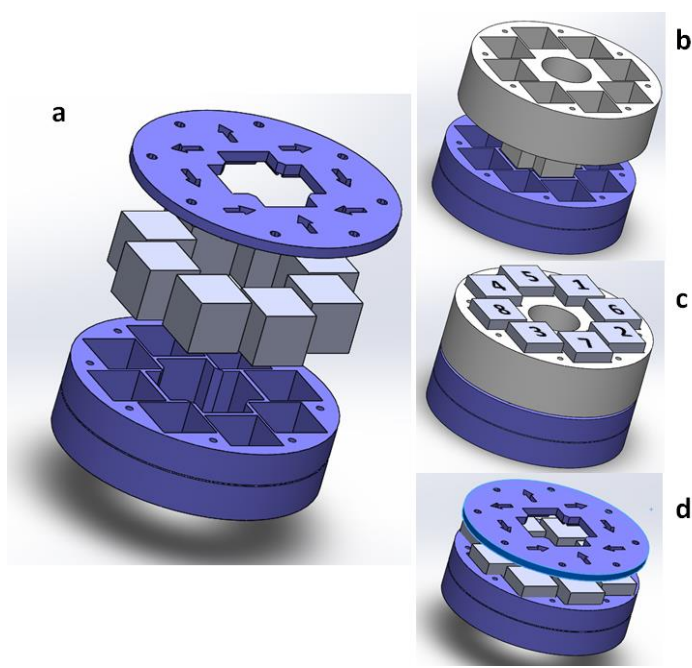 (PLA) assembly (see panel **a**). Stereolithography files (STL) for the three parts (frame, lid, and tool), ready to print in standard 3D printers, are available in the supplementary file *halbach.zip*. This is an extremely cost-effective setup to generate a magnetic field that is strong and homogeneous enough to align common thermotropic mesogens. The strength of the magnetic field, which is highest at the center of the magnet array, can be adjusted by modifying the vertical position of the sample with respect to the magnets. In order to facilitate the assembly, a custom tool is also recommended (gray part in panels **b** and **c**). The suggested assembly order of the magnets is labelled on panel **c**. The orientation of the magnetic moment of each magnet is marked both at the bottom of the assembly frame and on the lid cover. Each magnet should be firmly pushed to the bottom of the assembly frame. Additional magnets should be approached, one at a time, to its mounting hole from below the mounting frame to prevent their attraction to already mounted magnets. Once the eight magnets are assembled, they are kept in place by their mutual repulsion. It is then safe to remove the assembly tool and to put the lid cover, which may be fastened with M4 screws. **Note**: hole threading for fixing lid screws, or any additional holes needed to incorporate the magnet array into an optomechanical system must be performed **prior** to magnet assembly. *Warning: these magnets are dangerous, since they attract to ferromagnetic materials or to other magnets with unusual strength. Great care must be taken during assembly*.

Samples are held inside a thermostatic oven build with Thorlabs SM1 tube components and tape heater, and controlled with a Thorlabs TC200 controller.

## 1.5. Characterization

Routine observations of the active nematic are performed by means of conventional epifluorescence microscopy. We use a custom-made inverted microscope with a halogen light source and a Cy5 filter set (Edmund Optics). Image acquisition is performed with a QImaging ExiBlue CCD cooled camera operated with ImageJ µ-Manager open-source software.

For sharper imaging of the interfacial region, we employ laser-scanning confocal microscopy with a Leica TCS SP2 equipped with a photomultiplier as detector and a HeNe-633nm Laser as light source. We perform confocal acquisition both in fluorescence and reflection modes. While fluorescence confocal microscopy optimizes the signal/noise ratio for improved imaging of the interfacial material, we find that reflection confocal micrograph is optimal for image velocimetry of the active nematic due to the enhanced acquisition rate (Figs. S2 and S3). Moreover, this technique can be employed with label free active nematic, thus significantly simplifying sample preparation, reducing material costs, and, more importantly, eliminating extraneous moieties that might alter the way kinesin motors walk along the microtubules. Finally, fluorescence and reflection modes can be employed to simultaneously visualize both the active nematic and the passive liquid crystal (videos S2 and S3).

Tracer-free velocimetry analysis of the active nematic is performed with a public domain particle image velocimetry (PIV) program implemented as an ImageJ plugin (*8*). Further analysis of raw ImageJ output data is performed with custom-written MatLab code.

## 2. **Supplementary figures**

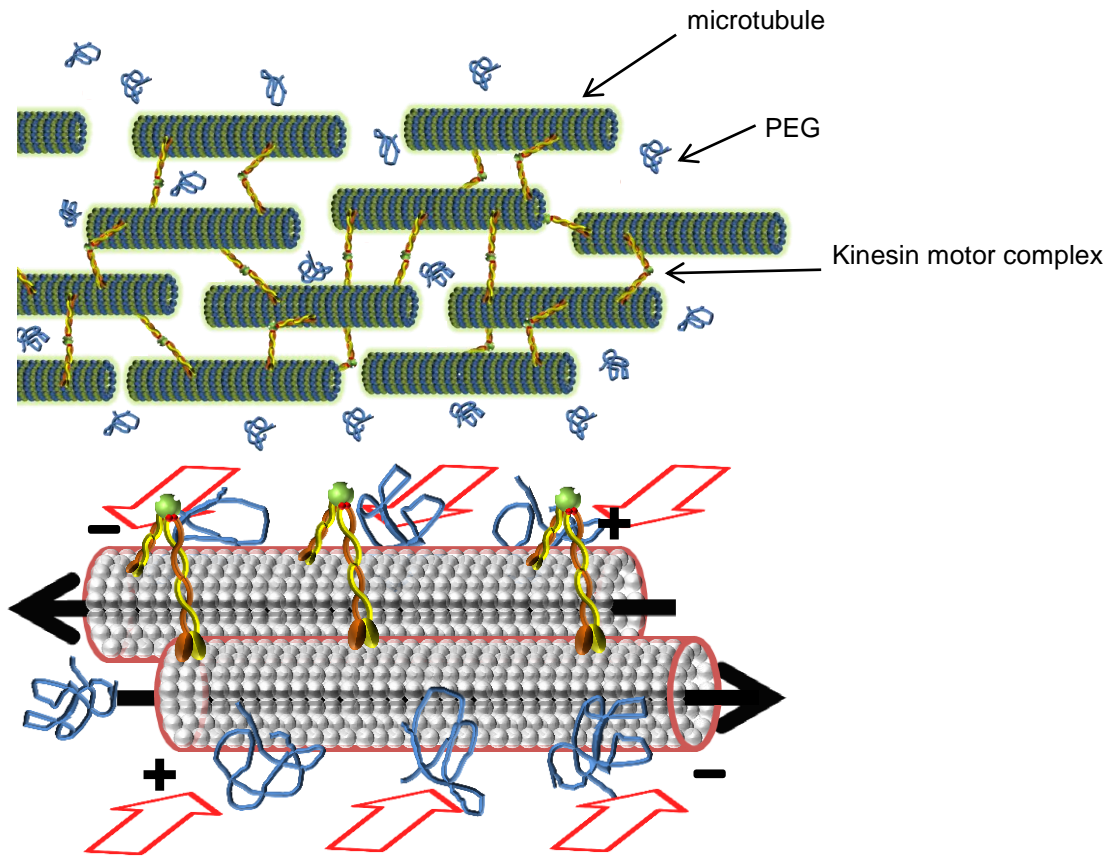

**Figure S1: Scheme of the active gel and its components.** Bundles of microtubules are formed via depletion interaction, which is triggered by the presence of polyethylene glycol (PEG) in the suspension, as illustrated by the red arrows. Within bundles, microtubules are cross-linked by ATP-activated kinesin motor clusters, which induce interfilament sliding. The local forces exerted by the protein complexes result in synchronized collective motility when adjacent MTs are oppositely aligned.

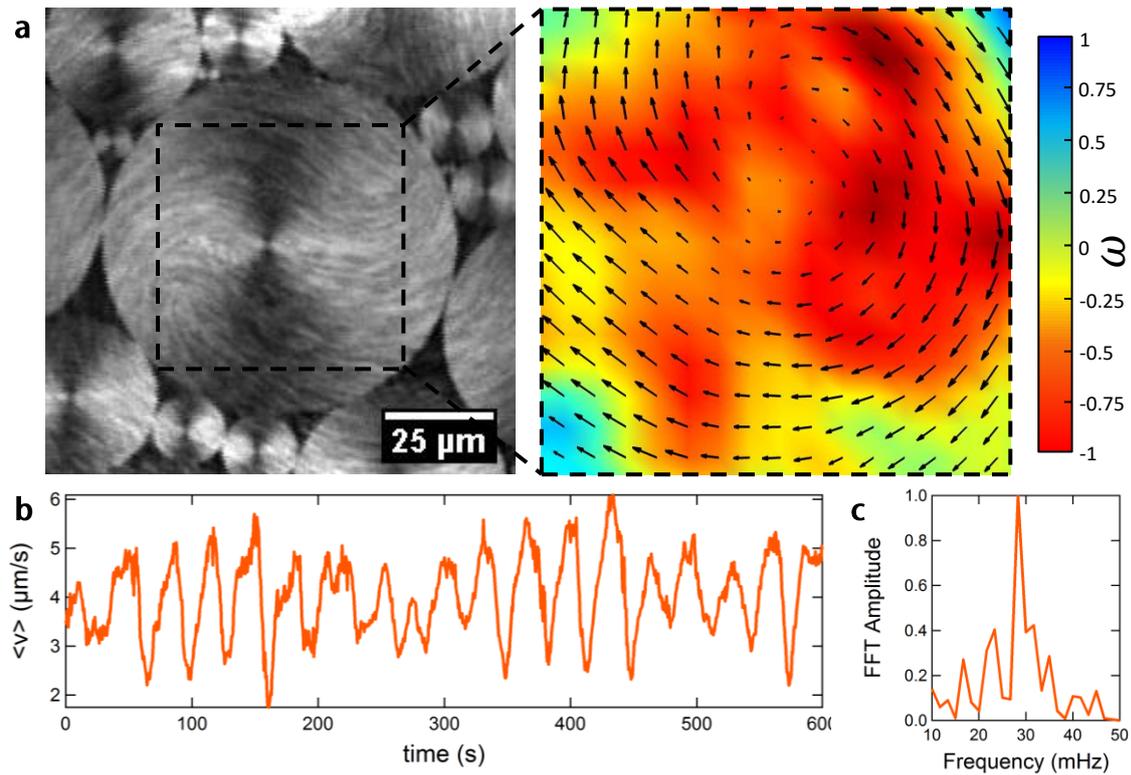

**Figure S2: Velocity and vorticity fields in an active nematic vortex.** (**a**) Reflection confocal microscopy permits the simultaneous observation of the structured LC and the active nematic (left) Spatial and temporal resolution is high enough to perform reliable image velocimetry of the active nematic. Output data is used to map the velocity field (right, black arrows) and calculate the normalized vorticity ($\omega = \vec{\nabla} \times \vec{v}$, right, colour map) at each point. (**b**) The average speed, <v>, oscillates at a characteristic frequency of 28 mHz. (**c**) Normalized Fast Fourier transform (FFT) of <v>.

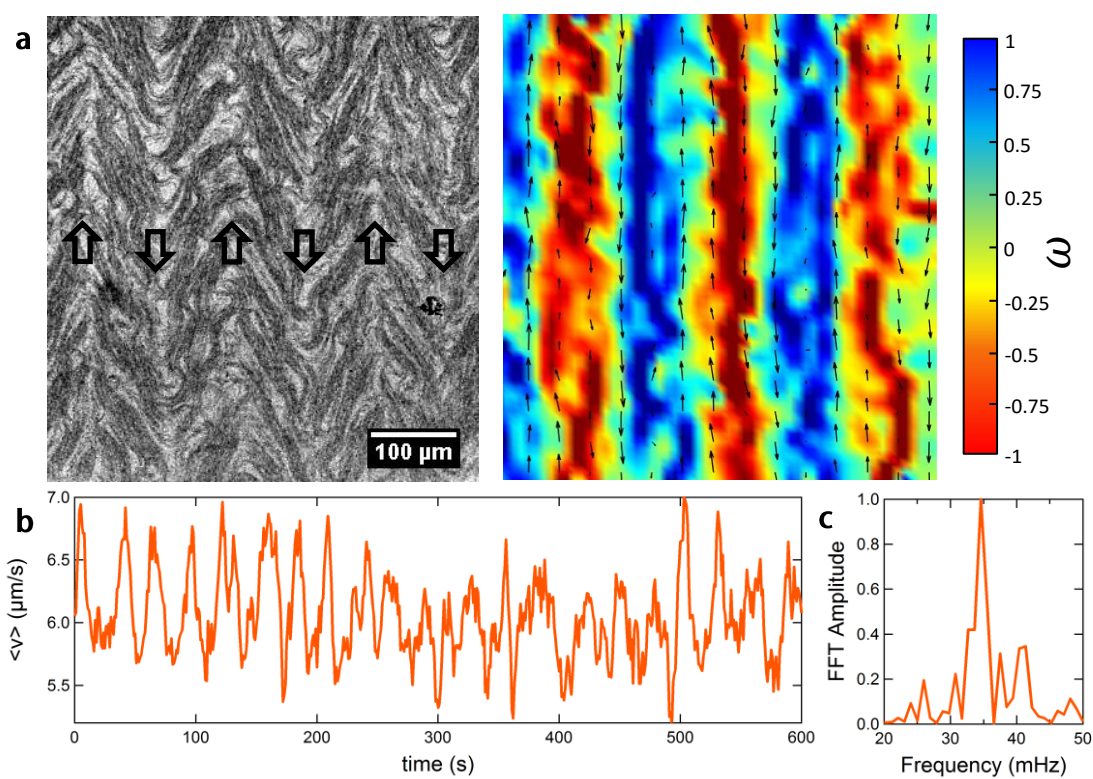

**Figure S3: Velocity and vorticity fields of the aligned active nematic.** (**a**) Reflection confocal micrograph (left). The arrowheads indicate the direction of defect motion (defects appear brighter than filaments with this observation technique). Corresponding velocity vector plot (right) is indicated with black arrows, which are superimposed to the normalized vorticity field ($\omega$, colour map). Note that vorticity maxima (in absolute value) are located on quasi-static and highly sheared regions (bundle stripes). Vorticity vanishes in fast moving regions (defect lanes). (**b**) Average speed oscillates in time due to the periodic disruption of the unstable aligned regions. (**c**) Normalized Fast Fourier transform (FFT) of <v>. The peak at 35 mHz is associated with the periodic disruption of the aligned active nematic, a consequence of the emergence of the bend instability inherent to its extensile nature.

### 3. Supplementary videos

**NOTE: Video files are available at www.ub.edu/socsam/AGSmA**

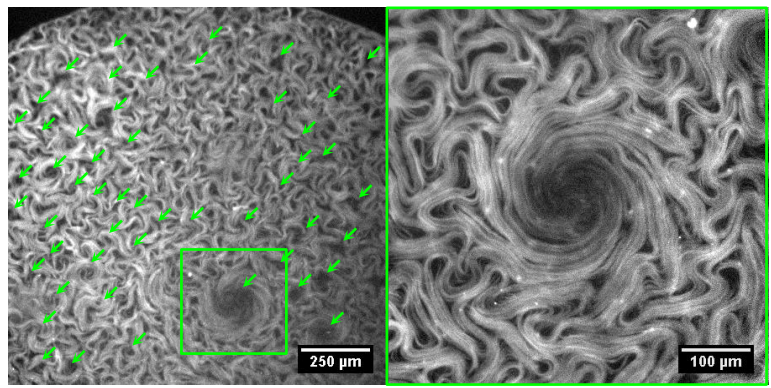

**Video S1: Fluorescence Micrographs of 2D Active Nematic Vortices.** In contact with toroidal focal conic domains, the active nematic organizes in vortices, led by the circular patterns in the oil phase. Green arrows (left) point to the location of vortexes. The square box frames the magnified area around the largest vortex (right).

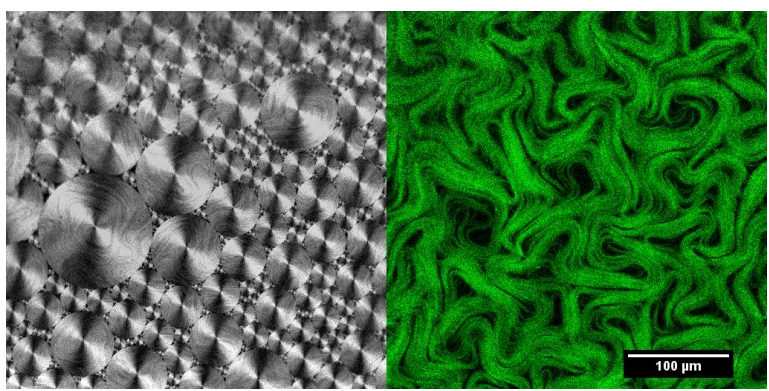

**Video S2: Transition from Vortices to Chaotic 2D Active nematic.** Acquisition with dual-channel confocal microscopy, simultaneously in reflection (left) and fluorescence (right) modes, evidences the influence of the interfacial patterning of the oil phase on the active morphology and dynamics of the active aqueous phase. In this video, 8CB melts from the smectic-A (showing toroidal focal conic domains at 25ºC) to the nematic phase (35ºC), causing remarkable changes in the active nematic phase. In brief, disappearance of the ordered pattern in the passive phase leads to the disaggregation of the vortices in the active phase. The number of defects in the active nematic decreases due to the reduction of the average shear viscosity.

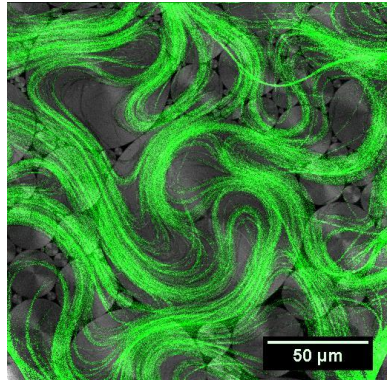

**Video S3: Preservation of the topological charge inside active vortices.** The video contains superimposed images of fluorescence and reflection confocal microscopy centered on a large active nematic vortex. The active nematic (green) is forced to follow the concentric direction of the toroidal focal conic domains in the oil phase (grayscale), with an accumulated defect charge constrained by the topology of a full circle (S=+1). The total number of defects varies but this topological constraint is satisfied at all times. Notice that small focal conics do not trigger the formation of vortices.

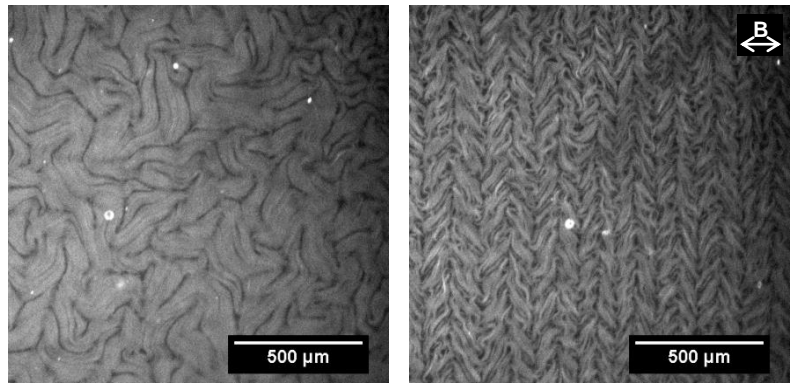

**VideoS4: Unidirectional alignment of the active nematic.** The normally chaotic active nematic dynamics is indirectly organized into unidirectional patterns by the action of a magnetic field. Transition from the nematic (35 ºC) to the smectic-A (25 ºC) phase of the passive liquid crystal 8CB, leads to the formation of a *bookshelf* configuration of the mesogen molecules under the influence of a homogeneous magnetic field (4 kG). The interface presents now strong uniaxial anisotropy of the shear stress, which forces the active nematic underneath. The biofilaments assemble now in stripes surrounded by defect lanes, which organize antiparallel flows along the direction of the less viscous axis, perpendicular to the magnetic field (B).

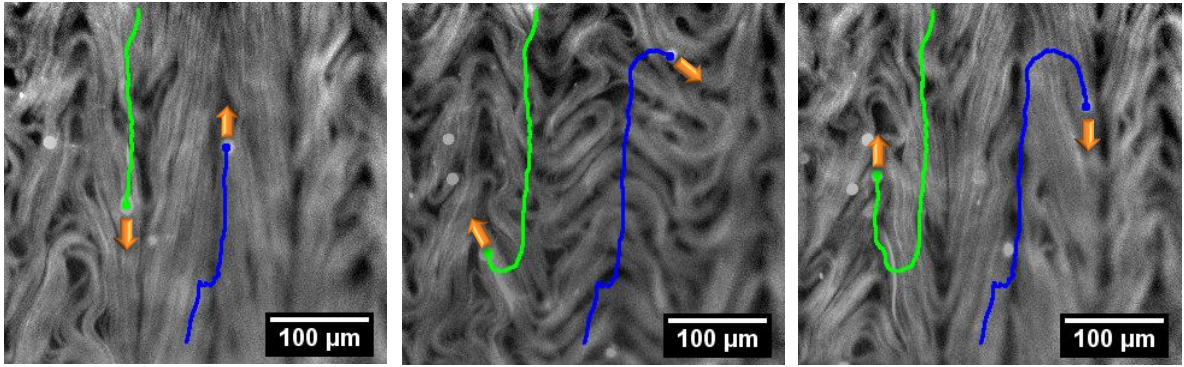

**Video S5: Aligned active nematic generates opposite unidirectional flows.** Passive tracer particles (PEGylated Polystyrene beads, 12 µm, Micromod, 08-56-124) are dragged along the defect lanes in the same direction but in opposite ways (left). Particle trajectories are highlighted in green and blue. Arrowheads indicate the direction of motion. When new defects proliferate (center), particles deviate from the original rectilinear trajectories and incorporate into preexisting adjacent lanes (right), thus reversing their direction. This corroborates the flow pattern sketched in Fig.4.

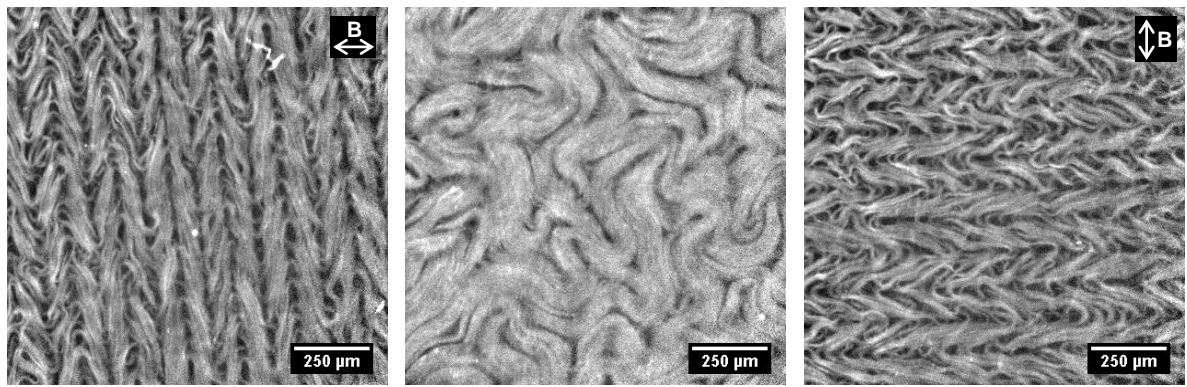

**Video S6: The aligned active nematic direction is *in situ*-rotated.** The aligned active nematic spontaneously disorders upon melting 8CB into the nematic phase at 35ºC. We subsequently rotate the magnetic field by 90º, and reverse the phase transition back into the smectic-A phase. The active nematic reestablishes the aligned state following the new easy axis, at 90º from the initial configuration.

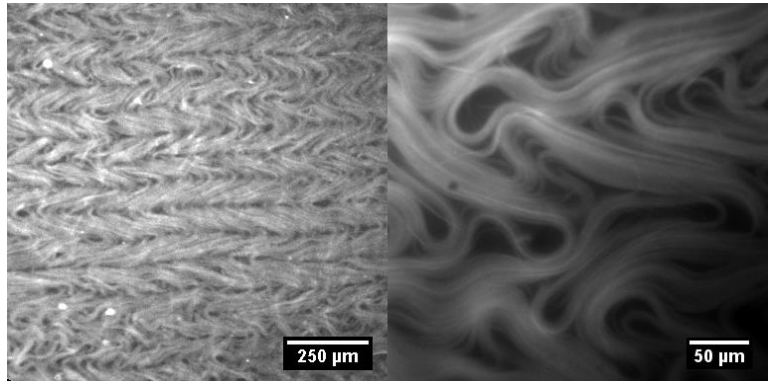

**Video S7: Periodic disruption of the aligned active nematic.** Bend instability periodically appears in the aligned stripes between defect lanes. This mechanism releases the extensile stresses accumulated in the active nematic while disrupting the directional dynamics of the active phase. The instability gives rise to new defects that will either annihilate or incorporate into preexisting defect lanes.